\documentclass[aps,showpacs,prx,amssymb,amsmath,twocolumn]{revtex4-2}
\usepackage{graphicx}
\usepackage{amssymb}
\usepackage{amsmath}
\usepackage{amsthm}
\usepackage{bm}
\usepackage{xcolor} 
\usepackage{array}
\usepackage{multirow}
\usepackage{tabularx}
\usepackage{booktabs}
\usepackage{textcomp}
\usepackage{mathtools}
\usepackage{hyperref}
\usepackage{qcircuit}
\usepackage{bbm}
\usepackage{cancel}
\usepackage{comment}
\usepackage{physics}
\usepackage{romannum}
\AtBeginDocument{\pagenumbering{arabic}}

\begin{document}

\title{Fast and reliable atom transport by optical tweezers} 
\author{Sunhwa Hwang$^{1}$, Hansub Hwang$^{1}$, Kangjin Kim$^{1}$, Andrew Byun$^{1}$, Seokho Jeong$^{1}$, Maynardo Pratama Soegianto$^{1,2}$, and Jaewook Ahn$^{1}$} 
\email{jwahn@kaist.ac.kr}
\address{$^{1}$Department of Physics, KAIST, Daejeon 34141, Republic of Korea}
\address{$^{2}$Department of Physics, Institut Teknologi Bandung, Jawa Barat 40132, Indonesia}  
\date{\today}

\begin{abstract} \noindent
Movable single atoms have drawn significant attention for their potentials as flying quantum memory in non-local, dynamic quantum computing architectures. However, when dynamic optical tweezers are employed to control atoms opto-mechanically, conventional methods such as adiabatic controls and constant jerk controls are either inherently slow or induce mechanical heating, leading to atom loss over long distances or at high speeds. To address these challenges, we explore the method known as shortcuts to adiabaticity (STA) as an efficient alternative for fast and reliable atom transport control. We present a series of proof-of-concept experiments demonstrating that STA-based optical tweezer trajectories can achieve both rapid and reliable single-atom transport. These experiments include moving atoms between two locations, adjusting speeds en route, and navigating curved trajectories. Our results indicate that atoms can be transported with a constant acceleration on average over distances that is only limited by trap lifetime, while effectively suppressing vibrational heating. This makes STA methods particularly well-suited for long-distance atom transport, potentially spanning distances over centimeter scales, such as between quantum information devices.
\end{abstract}
\maketitle

\noindent
There is a growing interest in neutral-atom quantum computing research~\cite{SaffmanMolmer2010_Review,Browaeys2020_Review,NorciaPRX2023,Saffman2022_qubitsEXP,Lukin2021_256qubits,LukinNat_movingEntangle2022,Bluvstein2024,KimSD2024}, largely due to the scaling potential of these systems, as demonstrated by their ability to trap thousands of atoms as individual qubits~\cite{Endres2023,NorciaArXiv2024}. At the same time, a significant challenge is emerging around effectively controlling and, in particular, moving these atoms spatially. As neutral-atom quantum devices are expected to partition space into distinct zones for operations like entanglement, storage, and readout, atoms are to be transferred between these zones~\cite{LukinNat_movingEntangle2022,NorciaPRX2023,Bluvstein2024}. As the system size increases, meaning the number of atoms and the area they occupy expand, the distance and time required for these transfers will also grow, raising the risk of disrupting their internal states. 

The traditional approach is the ``adiabatic" process~\cite{LewisJMP1969,PolkovikovNP2008,GueryOdelin2019_STAreview}, which preserves the quantum state by slowly evolving the system along the instantaneous eigenstates of a time-dependent Hamiltonian. Efficient atom transport within quantum information devices, traditionally requiring substantial time to prevent heating and preserve quantum states, has become increasingly important~\cite{PRA_Muga2011,C_Tuchendler_Energy_distribution, Sylvain2018,Lahaye_Traplifetime2021}. However, accelerating this process introduces side effects, referred to as the ``diabatic" process~\cite{GueryOdelin2019_STAreview,Couvert2008_MOT,HickmanPRA2020,HwangOptica2023}. Despite its effectiveness in maintaining the system's internal state, the inherent slowness of adiabatic operations can pose challenges when faced with practical time constraints in quantum computing operations.  

The method  of the shortcuts to adiabaticity (STA) in quantum mechanics offers a faster alternative to the adiabatic process, achieving the same outcomes of adiabatic processes without typical time constraints, making them suitable for time-sensitive scenarios~\cite{GueryOdelin2019_STAreview,Wang2018:QuatumGate_CD,Julia2012:ManyBodyModel_IIE,Kaufmann2018:TrappedIon_IIE,Ness2018:AtomicCloud_IIE}. We focus on fast and reliable single-atom transport by controlling optical tweezers along STA-based trajectories. Optical tweezers~\cite{AshkinPRL1970_tweezer, Schlosser2001_singleatomtweezer, GrimmAAMO2000, D_Alt2003} have been useful for manipulation of single atoms, which is essential for reconfiguring atom arrays~\cite{Nogrette2014_2Dtweezer, LeeOE2016_3dslm,LeePRA2017, Kim2016_2Drearrangement, BarredoSci2016_2daod} and facilitating qubit entanglements~\cite{Saffman2009_RydBlock, GaetanNP2009} in neutral atom quantum computing. In this paper, we will first verify the effectiveness of the STA method in an optical tweezer system and define its experimental limits. We will then conduct experiments to evaluate its applicability for generalized curved trajectories, in contrast to simple straight paths. Finally, we will discuss the potential advantages of this approach for long-distance transport, comparing it with other transportation methods. 

\vspace{0.5cm}
\noindent
{\bf STA-based single-atom transport} \\
\noindent
Our experiments utilize a two-dimensional (2D) atom array system described previously~\cite{
LeeOE2016_3dslm, Kim2016_2Drearrangement, LeePRA2017, HwangOptica2023, KimSD2024}. Laser-cooled rubidium atoms ($^{87}$Rb) were trapped in static tweezers generated by a 2D spatial light modulator (ODPDM-512 by Meadowlark Optics) and manipulated with dynamic tweezers controlled by a pair of acousto-optic deflectors (DTSXY-400-820 by AA Opto Electronics) and arbitrary waveform generators (AWG, M4i-6622-x8 by Spectrum Instrumentation and OPX+ by Quantum Machines). Both the static and dynamic optical tweezers were operated with a wavelength of 820~nm and a trap frequency of $\omega_0=\sqrt{2U_0/md^2} = 2\pi \times 90(10)$~kHz, where $U_0 = 0.8(2)$~mK is the optical potential depth, $m$ is the atom mass, and $d=0.73(7)$~$\mu$m is the harmonic-trap width~\cite{HwangOptica2023}. Atom detection was performed using fluorescence imaging of the $5S_{1/2}$-$5P_{3/2}$ transition, with an objective lens that had a numerical aperture of 0.5.

\begin{figure*}[htbp] 
\centering
\includegraphics[width=1.8\columnwidth]{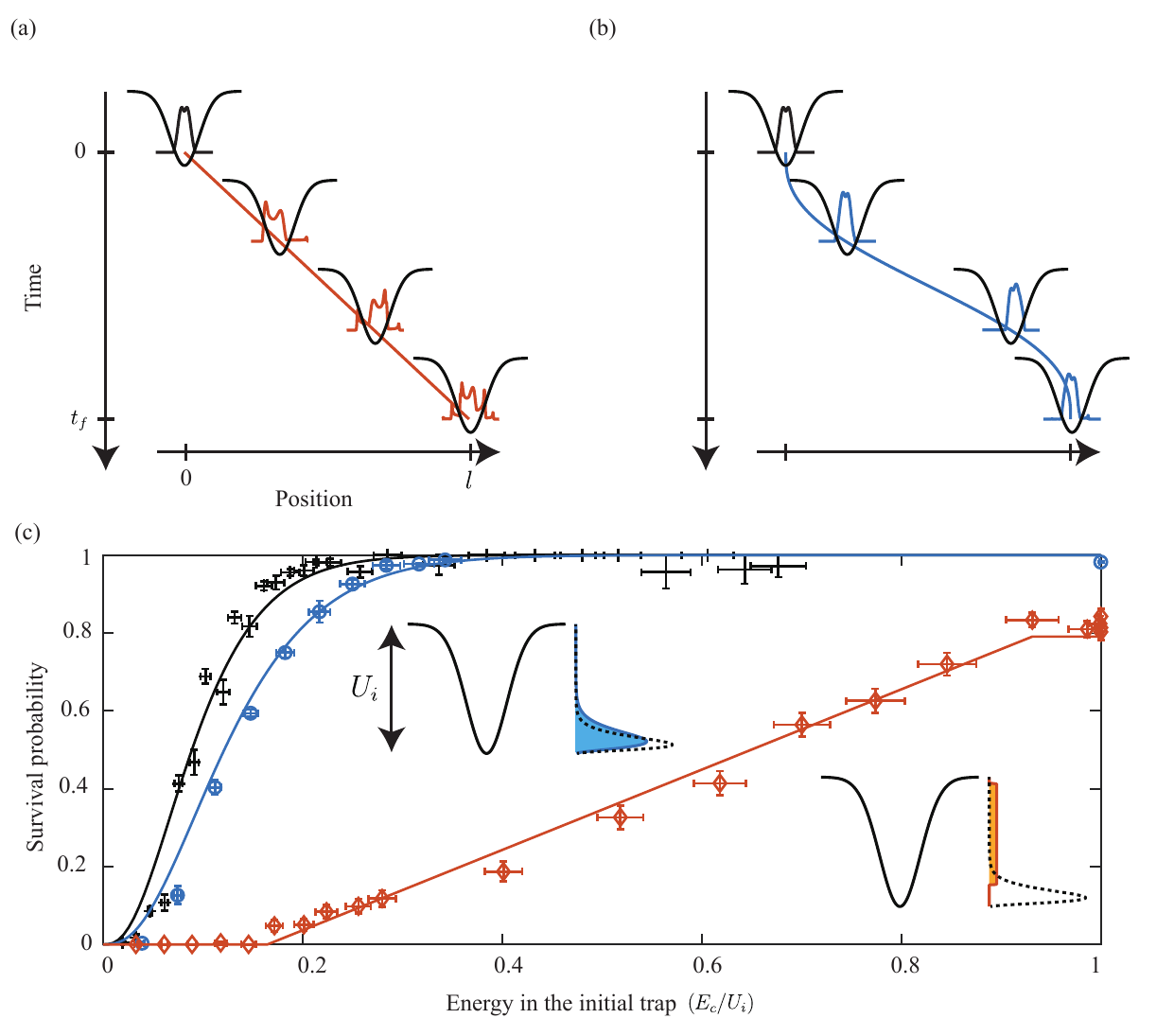}
\caption{STA-based atom transport trajectory compared with non-STA trajectories. (a,b) Schematic of transport trajectories of the optical tweezer with transportation duration $t_f=58.5$~$\mu$s and distance $l=12.6$~$\mu$s: (a) the STA-based solution in Eqs.~\eqref{atom} and \eqref{QC5}; (b) a corresponding constant velocity, non-STA trajectory. The black lines represent the optical potential of harmonically approximated optical tweezers. The shape of the atomic wave packet is well maintained through the STA path (blue), but it is deformed along the non-STA path (orange). (c) The survival probabilities of atoms after the transportation, measured as a function of the energy in the initial trap by varying the minimum trap depth $U_{\rm{min}}$ in adiabatic trap lowering~\cite{D_Alt2003, C_Tuchendler_Energy_distribution}. The black cross markers (blue circle markers, orange diamond markers) represent the survival probability after staying at the initial position (STA transportation, non-STA transportation) varying the cut-off atomic energy $E_c$ in unit of initial trap depth ($U_{\rm{i}}=0.8$~mK). The probability density distributions of atoms are correspondingly shown in the inset figures, with respect to the initial trap potential $U_{\rm{i}}$.}  \label{Fig1}
\end{figure*}

Figure~\ref{Fig1} illustrates the reliable performance of STA-based atom transport compared to a non-STA approach (constant-speed, ``adiabatic solution'' in a diabatic region). The optical tweezer path $x_o(t)$, shown as the blue line in Fig.~\ref{Fig1}(a), is designed to implement the STA-based atom transport along $x(t)$ using the invariant-based inverse engineering method~\cite{GueryOdelin2019_STAreview, PRA_Muga2011}. The atom's dynamics in the moving optical tweezer are governed by 
\begin{eqnarray} \label{eq1}
\ddot{x}+\omega_0^2 [x-x_o(t)]=0 \quad \mbox{for} \quad \abs{x-x_o(t)} <d,
\end{eqnarray}
and must satisfy the boundary conditions $x(0)=0$, $x(t_f)=l$, and $\dot{x}(0)=\ddot{x}(0)=\dot{x}(t_f)=\ddot{x}(t_f)=0$. When the tweezer's optical potential is approximated by a truncated harmonic potential~\cite{HwangOptica2023}, $U(x,t)={ U_0}(x-x_o-d)(x-x_o+d)/{d^2}$, the fifth-order polynomial solutions are given by
\begin{eqnarray} \label{atom}
\tilde x(t) &=& 10 \tilde{t}^3 - 15 \tilde t^4 + 6 \tilde t^5, \\
\tilde x_o(t)  &=& \tilde{x}(t) +  \frac{60}{\omega_0^2 t_f^2} \tilde t -\frac{180}{\omega_0^2 t_f^2}\tilde t^2 + \frac{120}{\omega_0^2 t_f^2} \tilde t^3.   \label{QC5}
\end{eqnarray}
in scaled position and time, $\tilde{x} = x/l$ and $\tilde{t} = t/t_f$.

The changes in atomic wavepackets during the STA and non-STA transport are shown in Figs.~\ref{Fig1}(a) and (b), respectively, by numerical simulations with transport conditions set as a distance of $l=12.6(3)$~$\mu$m, duration $t_f = 58.5(8)$~$\mu$s, and initial atomic temperature of $27(3)$~$\mu$K for both simulations. The STA transport is based on Eqs.~\eqref{atom} and \eqref{QC5}, successfully maintaining the initial atom conditions. The non-STA trajectory in Fig.~\ref{Fig1}(b) is a constant speed trajectory with the same average speed employed in a diabatic region, where the transport condition induces diabatic processes, i.e., $\Delta n \not\ll 1$,  with $\ket{n}$ representing vibrational energy state. While, at sufficiently slow transport speeds within an adiabatic region ($\Delta n \ll 1$), atoms following this uniform linear trajectory maintain their initial conditions, such as vibrational energy states and temperature, moving an atom along the non-STA trajectory in the diabatic region results in failure or deformation of the atom's initial state population, as shown in Fig.~\ref{Fig1}(b). 

Experimental results are shown in Fig.~\ref{Fig1}(c), where the atomic state distributions are measured after transportation, by adjusting the trap depth according to an adiabatic trap lowering sequence adapted from Refs.~\cite{D_Alt2003, C_Tuchendler_Energy_distribution}. The experimental data are shown for before (black crosses) and after transportation (STA with blue circles and non-STA with orange diamonds). The black and blue solid line represent the result of fitting the experimental data for before and after the STA transportation to the cumulative probability distribution function of the Maxwell-Boltzmann distribution, as $P(E_c) = 1 - [1 + \eta + \eta^2/2] e^{-\eta}$ with $\eta = E_c/k_BT$. In the left inset figure, the probability density distribution of atomic energy after STA transportation is shown by the blue solid line, compared to the black dashed line for no transportation. For an initial atom temperature of 27(3)~$\mu$K (black dashed line in the inset), the final atom temperature (blue solid line) of 36(4)~$\mu$K after STA transportation is measured, demonstrating that STA transportation effectively preserves the atomic energy distribution according to the Maxwell-Boltzmann distribution. In contrast, when the atom undergoes constant velocity movement along the non-STA path, it deviates from the Maxwell-Boltzmann distribution. Instead, the cumulative probability distribution (orange diamonds) closely follows a piecewise linear function, $P(x=E_c/U_i) =0$ ($0 < x < 0.165$), $1.03x - 0.17$ ($0.165 < x < 0.931$), $0.79$ ($0.931 < x < 1$), in agreement with expectations that the energy is distributed nearly uniformly among $\left\vert n \right\rangle$ states, where the average energy of the atom is $6.0(2) \times 10^{-27}$~J, which is equivalent to the mean energy of an atom at $0.29(1)$~mK that follows a Maxwell-Boltzmann distribution ($\bar{E} = {3}k_B T/2$). The probability density distribution of atomic energy after non-STA transportation, shown by the orange solid line in the right inset figure, reveals a significantly distorted, non-thermal energy distribution, suggesting substantial vibrational heating. Transport success probabilities are measured at $P=0.98(1)$ for the STA trajectory and 0.80(2) for the non-STA trajectory. The observed differences in atomic energy distributions and success probabilities indicate that the STA-based trajectory successfully transports the atom while preserving its initial states, whereas the non-STA approach falls into the diabatic transport condition.

\begin{figure*}[htbp] 
\centering
\includegraphics[width=2.0\columnwidth]{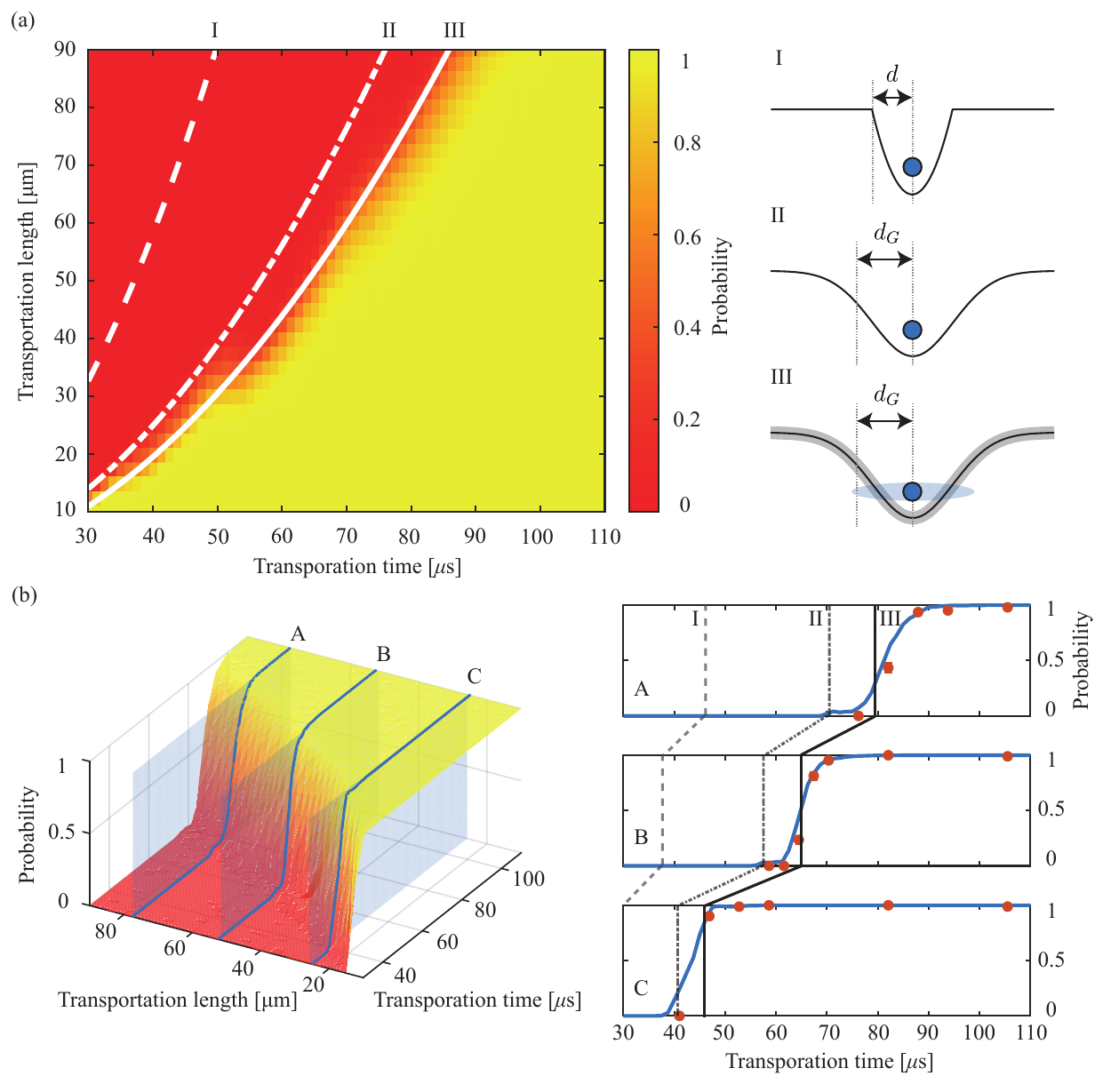}
\caption{Success probability $P(t_f, l)$  of STA-based transportation as a function of transportation distance $l$ and duration $t_f$. (a) Success probability within three different trap models, \Romannum{1} (a truncated harmonic trap), \Romannum{2} (a Gaussian distributed trap), and \Romannum{3} (a Gaussian distributed trap with atomic and trap fluctuations) on the right side. The background color contours represent the success probability of Model \Romannum{3} and the white dashed line (dash-dotted line, and solid line) denotes the boundary condition where atoms escape the optical tweezer for Model \Romannum{1} (\Romannum{2}, and \Romannum{3}). (b) Experimental results of STA-based transportation. As on the left side of the figures, the experiments are divided as three different groups, called as {A}, {B} and {C}, whose transportation distances are respectively 77.5(3), 51.7(3), and 25.2(3) ${\mu}m$. On the right-hand side, experimental results are plotted with orange circles for each of the three groups. The black dashed, dash-dotted, and solid vertical lines represent the boundary conditions of Model \Romannum{1}, \Romannum{2}, and \Romannum{3}, respectively. The blue solid line is the success probability of model \Romannum{3}, which is consistent with experimental results.} \label{Fig2}
\end{figure*}

\vspace{0.5cm}
\noindent
{\bf Optical tweezer models for effective speed limits} \\
\noindent
In the second experiment, we investigate the conditions for successful STA-based transport to obtain an effective speed limit of STA-based atom transport. Transport failure occurs when the atom escapes the tweezer, a scenario not accounted for in STA theory~\cite{PRA_Muga2011} that assumes an infinite harmonic trap. However, the optical tweezer has a finite trap potential and its distribution follows a Gaussian function rather than a harmonic one. To evaluate the effects of these discrepancies, we employ three different models of the optical tweezer, as depicted in Fig.~\ref{Fig2}(a), and correlate these findings with experimental results shown in Fig.\ref{Fig2}(b).

We first consider an idealized truncated harmonic trap, denoted by Model \Romannum{1} in Fig.~\ref{Fig2}(a), where a transportation failure occurs when the maximum displacement $\xi(t)={x}_o(t)-{x}(t)$ exceeds the trap width $d$ within the time interval $0 < t < t_f$, where the atomic trajectory and optical tweezer position are governed by Eqs.~\eqref{atom} and \eqref{QC5}, respectively. Consequently, allowed STA trajectories are confined by 
\begin{equation} \label{STA:limit}
l_{\rm \Romannum{1}} < \frac{\sqrt{3}}{5}\frac{U_0}{md} t_f^2,
\end{equation}
of which the boundary is illustrated with the white dashed line in Fig.~\ref{Fig2}(a). In the context of atomic state, this can be defined as the condition that the maximal energy state of the atom during STA transport will not exceed the trap energy level ($\Delta n \times \hbar \omega_0 < U_0$; see Methods for details).  So, the effective transportation speed is constrained by $\overline{v}_{\rm \Romannum{1}} < 0.17 \times a_{\rm max} t$, where $a_{\rm max}=U_0/md$~\cite{HickmanPRA2020,HwangOptica2023} is the maximal inertial acceleration. However, in real optical tweezer systems, Gaussian-distributed potential traps are wider than harmonic traps, as denoted by Model II in Fig.\ref{Fig2}(a). The gentler potential gradient in Gaussian traps produces a weaker net force on atoms, leading to larger displacements $\xi$ and a higher likelihood of transport failure along the same STA path. Taking this anharmonicity issue into account, we numerically calculated the atomic trajectory and displacement as $\xi_{\rm G}$. A Gaussian trap has a local extreme point at the trap radius $d_{\rm G}$, where the potential falls to $1/e^2$ of its maximum, especially in our system $d_{\rm G} = 1.0(1)~\mu$m. As the atom moves beyond $d_{\rm G}$, the restoring force diminishes rather than increases, preventing the atom from receiving enough force to return to the trap center and allowing it to escape. Thus, in Model \Romannum{2}, where we assume the atom escapes the trap if $\xi_{G}$ exceeds $d_G$, the boundary condition is represented by the dash-dotted line in Fig.~\ref{Fig2}(a) and the resulting effective transportation speed limit is approximated as $\overline{v}_{\rm \Romannum{2}}^{\rm max} \approx 0.429 \times \overline{v}_{\rm \Romannum{1}}^{\rm max}$. Model \Romannum{3} provides a more realistic trap representation by incorporating Monte Carlo simulations to account for experimental fluctuations, such as variations in atomic velocity, position distributions, and trap depth changes during optical tweezer controls. The probabilities of successful transportation are numerically estimated and depicted by the solid white line in Fig.~\ref{Fig2}(a). The successful transport probability, $P(t_f, l)$, is determined by recapturing atoms from a sample of 200, and is shown as the background color in Fig.~\ref{Fig2}(a). The boundary for successful transport is approximated by the solid white line, with the effective speed limit for transportation given by $\overline{v}_{\rm \Romannum{3}}^{\rm max} \approx 0.336 \times \overline{v}_{\rm \Romannum{1}}^{\rm max}$. 

Experimental results are presented in Fig.\ref{Fig2}(b) compared with the success probabilities predicted by the three atomic transport models. The experiments were carried out with three transportation distances of $l_{\rm A}=77.5(3)$~$\mu$m, $l_{\rm B}=51.7(3)$~$\mu$m, and $l_{\rm C}=25.2(3)$~$\mu$m, all under the same trap potential condition of $U = 0.8(2)$~mK, illustrating how they vary with different transportation times. The black dashed, dash-dotted, and solid lines represent the boundary conditions for the three models. The blue solid curves show the success probabilities for Model \Romannum{3} across different transport times, closely matching the experimental data.

\vspace{0.5cm}
\noindent
{\bf Concatenated and curved STA trajectories} \\
\noindent
We now explore the feasibility of generating general STA trajectories, including curved ones. Computing a path with a general shape involves defining specific boundary conditions for the trajectory and solving the relevant equations for each new shape. However, rather than recalculating solutions for each new path, we can break down a general trajectory into segments consisting of straight and rotational paths, with finite initial and final velocities, and then concatenate these segments. This approach requires two key validations: first, ensuring that concatenated STA paths preserve the initial atomic state after transport, and second, confirming that the generalized rotational path also maintains the initial state. Our next experiments will therefore  test whether combined paths remain valid as STA trajectories for atomic transport. 

\begin{figure}[bthp] 
\centering
\includegraphics[width=1.0\columnwidth]{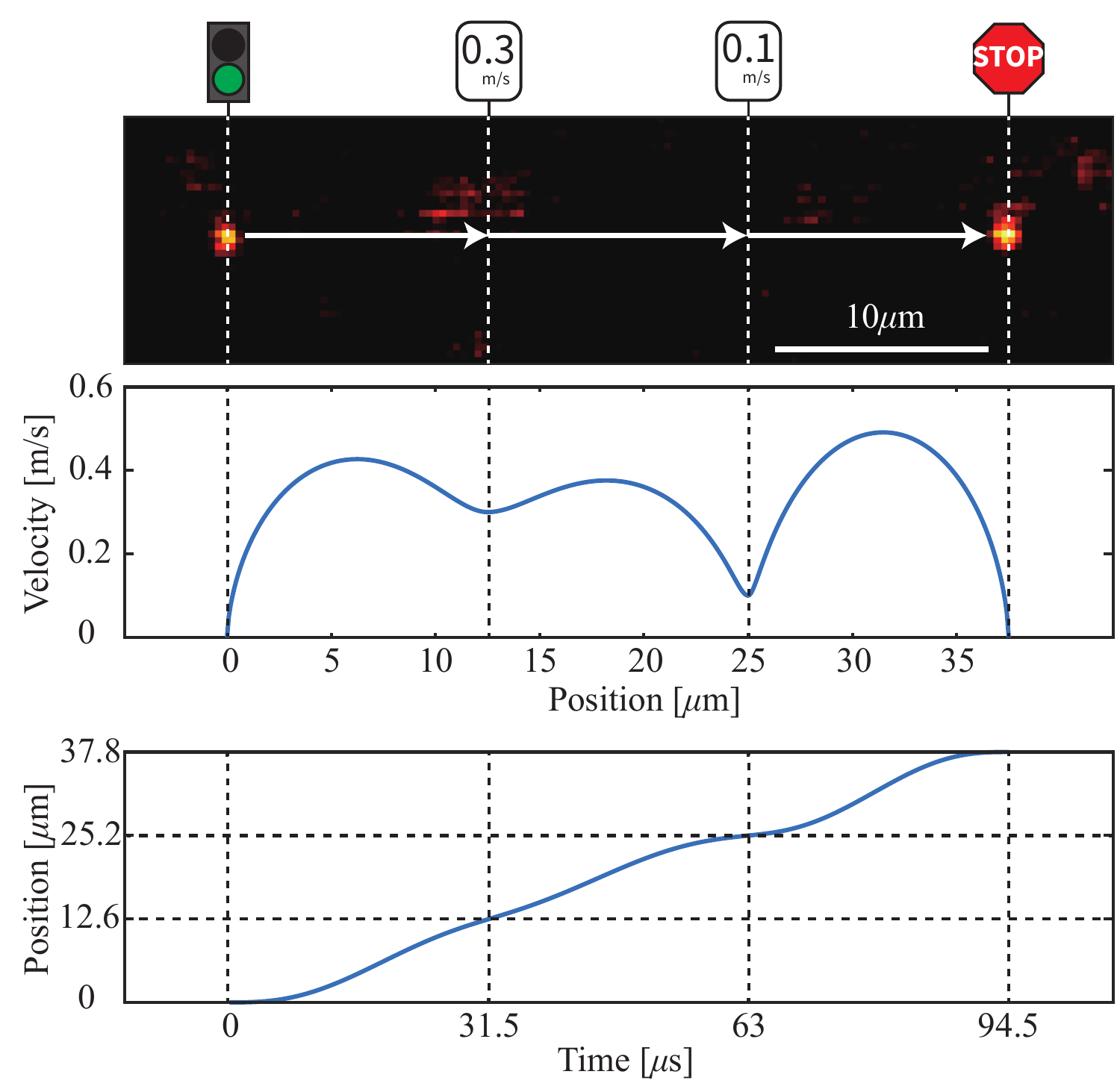}
\caption{A concatenation of three STA paths: An atom is guided along three STA segments with the same duration $t_f=31.5(8)~\mu$s  and distance $l=12.6(3)~\mu$m but with different initial and final velocities, (\romannum{1}) $v_i=0~$m/s and $v_f=0.3~$m/s, (\romannum{2}) $v_i=0.3~$m/s and $v_f=0.1~$m/s, and (\romannum{3}) $v_i=0.1~$m/s, and $v_f=0~$m/s, respectively. The transportation success probability is measured to $P=0.99(1)$ and the final temperature  $T_f=15(3)~\mu$K, when the initial temperature is $T_f=12(2)~\mu$K.}  \label{Fig3}
\end{figure}

Figure~\ref{Fig3} presents an atom transport experiment along three concatenated STA paths with different boundary conditions. The atom's velocities are non-zero at the connection points between segments. Following a similar procedure to that used for Eq.~\eqref{atom}, the STA solution for nonzero, initial and final velocities, $v_i$ and $v_f$, is obtained as $x(t)=  v_i t + (10d -6v_i t_f-4v_f t_f) \tilde{t}^3 - (15d -8v_i t_f-7v_f t_f) \tilde{t}^4 + (6d -3v_i t_f-3v_f t_f) \tilde{t}^5$,
with all other conditions remaining the same (see Methods for details). Using this equation, we calculate three STA linear paths so that each of the segments is of the same travel distance $12.6(3)$~$\mu$m within same transportation time $31.5(8)$~$\mu$s, while each features different initial and final velocities.This transportation is performed outside of the adiabatic transport region, with a transport speed of $\bar{v} = 0.4$~m/s. Experimental result in Fig.~\ref{Fig3} shows that a high success probability of $P = 0.99(1)$ is achieved and the atomic temperature is maintained at 15(3)~$\mu$K, close to the initial temperature of 12(2)~$\mu$K, confirming that concatenating STA paths can indeed result in a valid STA path.

\begin{figure}[htb!] 
\centering
\includegraphics[width=1.0\columnwidth]{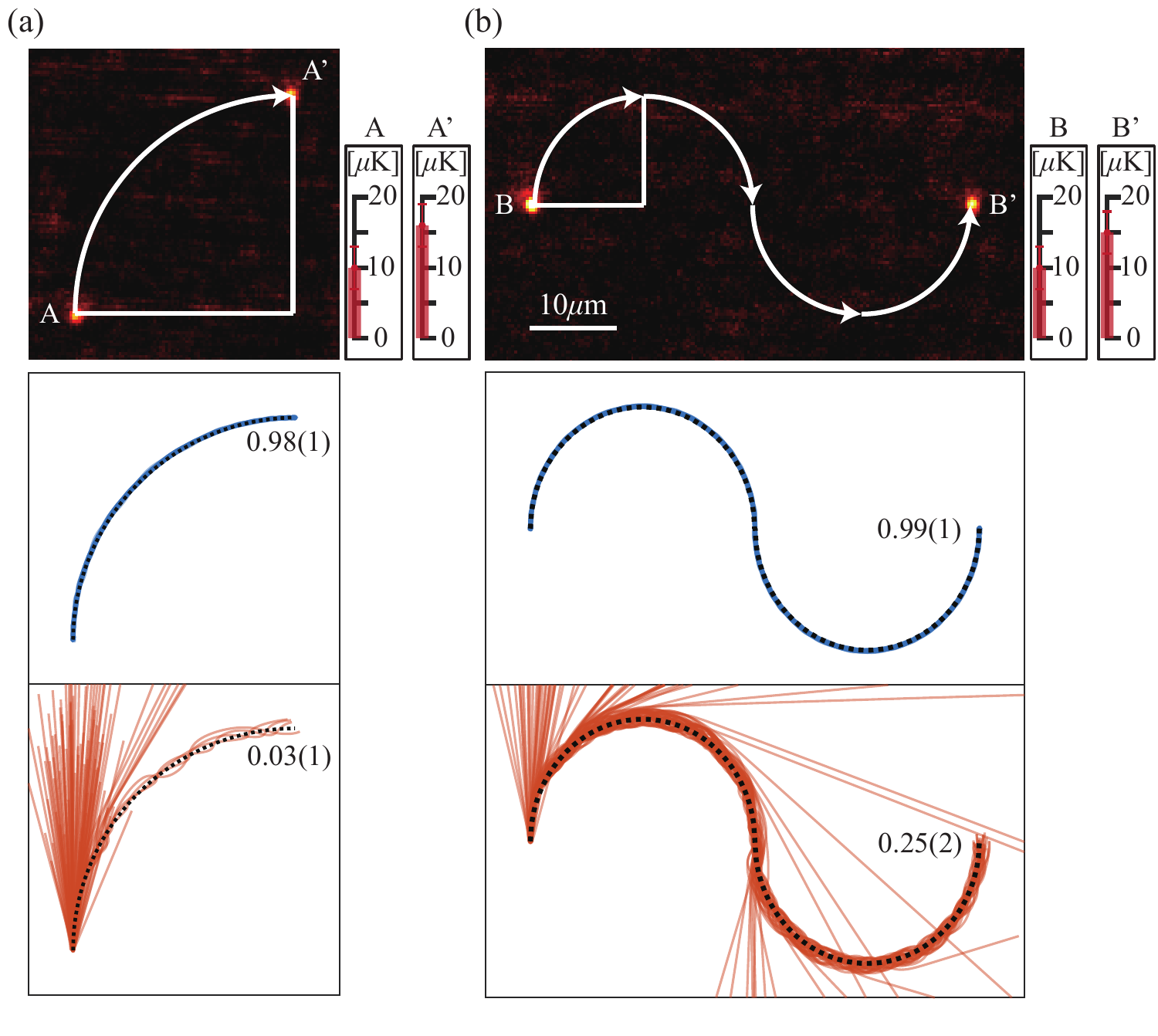}
\caption{Curved STA trajectories for atom transport: (a) A 90-degree circular STA trajectory. The top subgraph shows an illustrative  image of atoms before and after the rotational transportation. As a non-STA transportation, the constant angular velocity trajectory ($r(t) = R$, $\theta (t) = \theta_f t/t_f$) is utilized, satisfying the same transportation distance and duration. For the middle and bottom subgraphs, the blue and orange lines represent the atom trajectories during the STA transportation and non-STA transportation, respectively, and the black dotted line represents the trajectory of the trap center for each transportation. (b) An S-shaped STA trajectory. The S-shaped atom trajectory is designed by combining two STA-based semicircular paths, each rotating 180 degrees with the same radius $R=12.6(3)$~$\mu$m within the same duration $t_f=128.8(8)$~$\mu$s but in opposite directions. The final speed of the first semicircle and the initial speed of the second semicircle are $v_{\text{inter}} = 0.3~$m/s. As a non-STA transportation, two of the constant angular velocity trajectories with $r(t) = R$, $\theta (t) = \theta_f t/t_f$ are combined.}   \label{Fig4}
\end{figure}

Curved STA trajectories are tested and summarized in Fig.~\ref{Fig4}. STA-based solutions for rotational atomic transport can be classically simplified in polar coordinates, by leveraging the radial symmetry of the optical tweezers~\cite{2D_Muga2022}. For the general boundary conditions, $\theta(0)=\theta_{o}(0)=0$, $\theta(t_f) =\theta_{o}(t_f)=\theta_f$, $\dot{\theta}(0)=v_i/R$, and $\dot{\theta}(t_f)=v_f/R$, the STA-based rotational path with a fixed radius $R$ is obtained as $\theta(t) = ({v_it_f})\tilde{t}-({6v_i t_f+4v_f t_f} - 10 R \theta_f)\tilde{t}^3 + ({8v_i t_f+7v_f t_f}-15 R\theta_f )\tilde{t}^4 - ({3v_i t_f+3v_f t_f}-6 R\theta_f )\tilde{t}^5$, where $\tilde{\theta} = \theta/\theta_f$ and $\tilde{t} = t/t_f$  (see Methods for details). A 90-degree rotation is illustrated in Fig.~\ref{Fig4}(a), where $R = 25.2(3)$~$\mu$m, $v_i = v_f = 0~$m/s, $t_f = 93.7(8)$~${\mu}$s, and $\theta_f =\pi/2$. For this case of $v_i = v_f = 0$, the solution is given by $\tilde{\theta}(t) = 10 \tilde{t}^3 - 15 \tilde{t}^4 + 6 \tilde{t}^5$. When we compare the success probability of this STA rotational transport with that of a constant angular velocity optical tweezer, defined by $r(t) = R$ and $\dot{\theta} =\theta_f/t_f$, to see if the atoms maintain their initial state in the diabatic region, an experimental as-designed STA transport results in a higher success transport probability of $P=0.98(1)$ compared to $P=0.03(1)$ of the non-STA method. The temperature is well maintained at 16(3)~$\mu$K close to the initial temperature of 10(3)~$\mu$K. An `S'-shaped transportation path is also implemented by combining two semicircular paths based on the STA-based rotation solution, as illustrated in Fig.~\ref{Fig4}(b). The semicircular paths rotate 180 degrees in opposite directions with the same radius of $R=12.6(3)$~$\mu$m within duration of $t_f=128.8(8)$~$\mu$s. To maintain continuous motion without reducing speed to zero at the intersection, the final speed of the first semicircle is matched to the initial speed of the second semicircle, i.e., $R\dot{\theta_1}(t_f) = R\dot{\theta_2}(0) = 0.3$~m/s. For comparison, non-STA transport is tested with a constant average angular velocity $\dot{\theta} = {\pi}/t_f$. Model III simulations show that the STA atom trajectory, which is the middle subgraph in Fig.~\ref{Fig4}(b), is more convergent to the trap center than the non-STA trajectory in the bottom subgraph in Fig.\ref{Fig4}(b), consistent with experimental success probabilities of $P=0.99(1)$ for STA and $P=0.25(2)$ for non-STA transport. Additionally, the atom's temperature is well maintained in the STA transportation, with the final temperature of 15(3)~$\mu$K matching the initial temperature of 10(3)~$\mu$K. 

\begin{figure}[htbp]
\centering
\includegraphics[width=1.0\columnwidth]{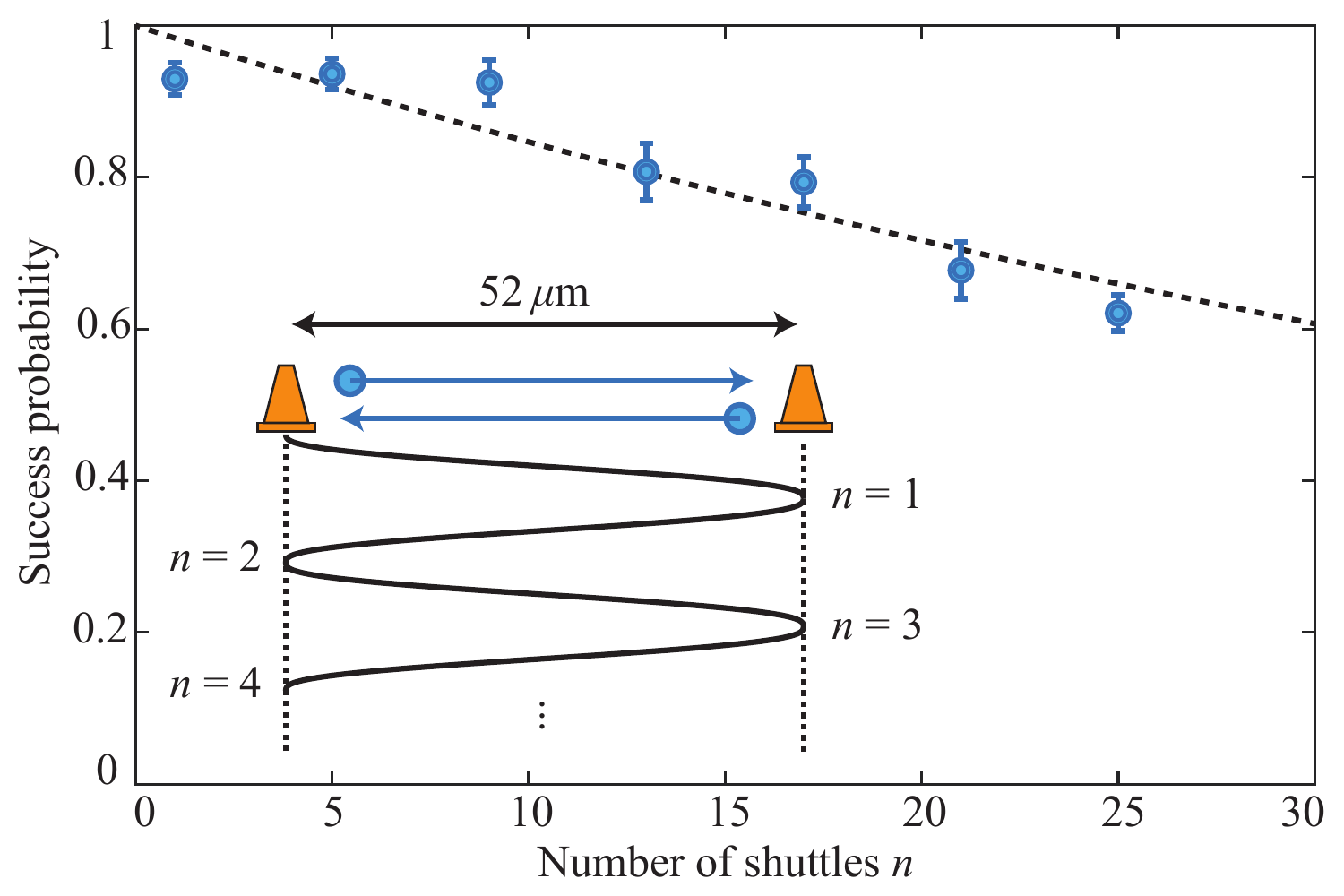}
\caption{Atomic shuttle run experiment: An atom is shuttled back and forth between two locations using an STA path with a duration of $t_f = 129.0(8)$~$\mu$s over a distance of $l=51.8(3)$~$\mu$m. Success probabilities are measured as a function of one-way run number $n$ for up to $n=25$ and fitted to $P_s(n)= 0.984(4)^n$.} \label{Fig5}
\end{figure}

\vspace{0.5cm}
\noindent
{\bf Atomic shuttle running} \\
\noindent
In the context of long-distance atom transport, we have conducted an experiment of repeatedly shuttling an atom between two positions, which we may term ``atomic shuttle running".  This test estimates the maximum transport distance achievable within the field of view, which is approximately 100~$\mu$m squares, limited by our EMCCD imaging device. As illustrated in the inset of Fig.~\ref{Fig5}, each one-way travel covers a distance of $l = 51.7(3)$~$\mu$m, with a runtime of $t_f = 129.0(8)$~$\mu$s. Success probabilities $P_s(n)$ are measured for various repetitions up to $n = 25$ and numerically fitted to
\begin{equation}
P_s(n) = {0.984(4)}^n.
\end{equation}
This result indicates that the STA-based atom trajectory is expected to span over one centimeter with a success probability exceeding $P_s(n=200) = 3.9\%$. In contrast, when using a constant-velocity trajectory under the same transport conditions, the success probability is significantly lower at $P_s(n=1) = 0.42(6)$, implying a transport distance of about 200~$\mu$m with 3.1\% success rate\textemdash50 times shorter than the STA approach.

\vspace{0.5cm}
\noindent
{\bf Discussion and conclusion} \\
\noindent
In the context of long-distance atom transportation, the STA approach offers notable advantages over the constant velocity (CV, $\tilde{x}_{\rm CV} (t) = \tilde{t}$ ) and constant jerk (CJ, $\tilde{x}_{\rm CJ}(t) = 3\tilde{t}^2-2\tilde{t}^3$, for example) trajectories that are currently employed in atomic transportation~\cite{Lahaye2020,LukinNat_movingEntangle2022}. This can be deduced based on the energy state changes that the atoms undergo along each transportation path. In the event that the optical tweezer is considered to be a quantum harmonic oscillator, the state of the atom is determined by the force it receives, that is to say, the amount of acceleration~\cite{Nieto_dN_1965}. If we assume that the atom travels a considerable distance in relation to the trap size and focus on the scenario where the transport is successful, we can disregard the discrepancy between the atom's trajectory and the trap's ($\xi_{\rm max} < d  \ll l$).  Based on these assumptions, the real-time acceleration of the atom is approximated as the one of the trap. From this, we can calculate the maximum distance the atom can successfully travel in a finite depth of the trap, by considering a situation where the maximum energy of the atom is equal to the energy of the trap (max$\left\vert \Delta n (t) \right\vert \cdot \hbar \omega_0 = U_0$). Therefore, the maximum transportable distances scale as (see Methods for details):
\begin{eqnarray}
l_{\rm max}=
\begin{cases}
\sqrt{ \frac{2 U_0}{m}}  \times t_f & \mbox{for CV path} \\
\frac{1}{3\sqrt{3}} \frac{U_0}{md} \times t_f^2 & \mbox{for CJ path } \\
\frac{\sqrt{3}}{5} \frac{U_0}{md} \times t_f^2 & \mbox{for STA path}.
\end{cases}
\end{eqnarray}
While both CJ and STA achieves $l_{\rm max}~\propto t_f^2$, a constant average acceleration motion, they differ in the degree of atom heating, which is dependent upon the final state of the atom, $\Delta n $, following movement, as illustrated below for each trajectory. 
\begin{eqnarray}
\Delta n \simeq
\begin{cases}
\frac{m l^2}{2\hbar \omega_0} \times t_f^{-2}  & \mbox{for CV} \\
\frac{36 m l^2}{\hbar \omega_0^3} \times t_f^{-4} & \mbox{for CJ }  \\
\frac{3600 m l^2}{\hbar \omega_0^5} \times t_f^{-6} & \mbox{for STA}.
\end{cases}
\end{eqnarray}
In the case of 100~$\mu$s, the final state increases of the atoms are 188, 125, and 13, respectively. It should be noted that, in contrast to CJ and CV, the maximum and final state changes of STA is not of the same order of magnitude over time. Consequently, the farther and longer the atom travels, the smaller the final temperature change for the same probability of success in STA.  As an illustration, if atoms traverse a distance of 1~ms, the potential maximum distance for CV, CJ, and STA are 391~$\mu$m, 21~mm, and 38~mm, respectively. Notably, the vibrational state changes of the atom diminishes from 13 to 0.1 for STA trajectory alone, in contrast to the other two paths, which have same increases for 100~$\mu$s.

Aside from technical limitations, the success of atom transportation is constrained by trap lifetime $\tau_{\rm trap}$, which defines the maximum duration that an optical tweezer can stably hold atoms. The state-of-the-art trap lifetime is currently 6000 seconds~\cite{Lahaye_Traplifetime2021}, allowing for 60 seconds of stable atom guidance with a probability over $0.99$. Within this timeframe, an atom could theoretically travel a distance of approximately $3.6 \times10^7$~m at a maximum atom velocity of $6.0\times10^5$ m/s. However, the limitation of the AOD specification~\cite{AODSpec} restricts the possible trap velocity ($v_{\rm trap, AOD} \le 66$ m/s), so the atom can expectedly travel a distance 80~mm through 2.3~ms, where the average velocity of transportation is about 35~m/s. Also, an additional technical constraint is the size of the camera windows, presently the largest one is approximately $20$~mm~\cite{QcmosSpec}, thus the maximum achievable average velocity could reach $14$~m/s. Given the constraints of our current experimental setup, particularly the EMCCD size (approximately 100~$\mu$m), the maximum transport speed in our experiments is estimated to be 0.88~m/s over a distance of $l = 78$~$\mu$m and a duration of $t_f = 87$ ~$\mu$s.

In conclusion, this study demonstrates that the shortcuts to adiabaticity (STA) method significantly improves the efficiency, precision, and range of atomic transport in optical tweezers, making STA particularly suitable for operating dynamic quantum information devices and other applications requiring precise atomic control. Both experimental and simulation results confirm that STA-based trajectories preserve the Maxwell-Boltzmann energy distribution and avoid vibrational wave packet deformations. 
The versatility of STA was further illustrated by successfully applications to complex curved trajectories, such as S-shaped paths, through the concatenation of multiple STA segments and implementing STA-based rotations, achieving both the high success rate and minimal temperature change. Overall, STA method offers significant advantages in optimizing experimental efficiency by enabling faster and reliable atom transport, extending the feasible transportation distance within practical trap operation times, thereby proving especially valuable for quantum information devices.

\vspace{0.5cm}
\noindent
\section*{Methods} 


\vspace{0.5cm}
\noindent
{\bf STA-based atom transport solutions} \\
\noindent
An optical tweezer path $x_o(t)$ is designed to implement STA-based atom transport along $x(t)$ using the invariant-based inverse engineering technique~\cite{GueryOdelin2019_STAreview, PRA_Muga2011}. The optical tweezer's potential is approximated as a truncated harmonic potential~\cite{HwangOptica2023}, given by
\begin{eqnarray} \label{U:har}
U(x,t)&=&\frac{ U_0}{d^2}[x-x_o(t)-d][x-x_o(t)+d],
\end{eqnarray}
where $U_0$ is the optical potential depth, $d$ is the width of the optical tweezer, $\omega_0=\sqrt{2U_0/md^2}$ is the trap frequency, and $m$ the atom mass. The atom's dynamics is then governed by 
\begin{eqnarray} \label{eq1}
\ddot{x}+\omega_0^2 [x-x_o(t)]=0 \quad \mbox{for} \quad \abs{x-x_o(t)} <d.
\end{eqnarray}
Using this harmonic approximation, the appropriate atom path that ensures its final state matches its initial state is described by (in scaled position and time, $\tilde{x} = x/l$ and $\tilde{t} = t/t_f$):
\begin{eqnarray} \label{1Dgeneral}
x(t)&=&  v_i t + (10d -6v_i t_f-4v_f t_f) \tilde{t}^3 \nonumber \\
&-& (15d -8v_i t_f-7v_f t_f) \tilde{t}^4 \nonumber \\
&+& (6d -3v_i t_f-3v_f t_f) \tilde{t}^5,
\end{eqnarray} 
using a polynomial ansatz to satisfy the boundary conditions,
\begin{subequations} \begin{eqnarray}  \label{bc}
x(0)=0, &~\dot{x}(0)=v_i,  &~\ddot{x}(0)=0, \label{bc1}\\ 
x(t_f)=l, &~\dot{x}(t_f)=v_f, &~\ddot{x}(t_f)=0,  \label{bc2}
\end{eqnarray}
\end{subequations}
to transport the atom from $x=0$ to $x=l$ over the time interval from $t=0$ to $t=t_f$. The optical tweezer trajectory, $x_o(t)$, facilitating the atom transport path in Eq.~\eqref{1Dgeneral}, is then obtained from the equation of motion in Eq.~\eqref{eq1}. For the specific case where the atom's initial and final velocities are both zero, $v_i = v_f = 0$, the atom and optical tweezer trajectories are simplified as follows:
\begin{eqnarray} \label{atomb}
\tilde x(t) &=& 10 \tilde{t}^3 - 15 \tilde t^4 + 6 \tilde t^5, \\
\tilde x_o(t)  &=& \tilde{x}(t) +  \frac{60}{\omega_0^2 t_f^2} \tilde t -\frac{180}{\omega_0^2 t_f^2}\tilde t^2 + \frac{120}{\omega_0^2 t_f^2} \tilde t^3.   \label{QC5b}
\end{eqnarray}

Furthermore, two-dimensional scenarios, such as rotational paths with a fixed radius $R$, can be addressed using the inverse engineering method~\cite{2D_Muga2022}. Solutions for curved paths in two dimensions can serve as a basis for designing more complex, multi-shaped paths. Unlike the computationally challenging general solution~\cite{2D_Muga2022}, STA-based solutions for rotating atomic transport can be classically simplified in polar coordinates ($\tilde{\theta} = \theta/\theta_f$ and $\tilde{t} = t/t_f$), by leveraging the radial symmetry of the optical tweezers, as follows:
\begin{eqnarray} \label{2Dgeneral}
\theta(t) &=& \left(\frac{v_it_f}{R}\right)\tilde{t}-\left(\frac{6v_i t_f+4v_f t_f}{R} - 10 \theta_f\right)\tilde{t}^3 \nonumber \\
&+& \left(\frac{8v_i t_f+7v_f t_f}{R}-15 \theta_f \right)\tilde{t}^4 \nonumber \\
&-& \left(\frac{3v_i t_f+3v_f t_f}{R}-6 \theta_f \right)\tilde{t}^5.
\end{eqnarray}
These simplified solutions need to satisfy the boundary conditions:
\begin{subequations} \begin{eqnarray} \label{2Dbc}
&\theta(0)=\theta_{o}(0)=0, & \quad  \dot{\theta}(0)=v_i/R, \\
& \theta(t_f) =\theta_{o}(t_f)=\theta_f, & \quad  \dot{\theta}(t_f)=v_f/R.
\end{eqnarray} 
\end{subequations}
For the special case where $v_i = v_f = 0$, the solution is given by $\tilde{\theta}(t) = 10 \tilde{t}^3 - 15 \tilde{t}^4 + 6 \tilde{t}^5$. In Cartesian coordinates, the optical tweezers and the moving atom satisfy the equation of motion given by  $(x_o,y_o) = (x,y) + (\ddot{x}, \ddot{y})/{\omega_0^2}$.

\vspace{5cm}
\noindent \\
\noindent
{\bf Atomic energy distribution measurement} \\
\noindent
The adiabatic trap lowering is an experimental method to measure the energy distribution of atom in the optical tweezer~\cite{D_Alt2003}. This sequence is consisted of three steps:  (1) slowly decreasing trap depth to minimum trap depth $U_{\rm{min}}$; (2) holding trap depth at $ U_{\rm{min}}$ for sufficient time;  and (3) increasing trap depth slowly to initial trap depth $U_{\rm{i}}$. During the first and last steps, the slow operation of optical tweezers is kind of the adiabatic process, so the atomic vibrational mode is maintained. At the middle step, since the atoms having higher energy escape the optical tweezers, and the remianing atoms after this whole process have lower energy than cut-off energy $E_{\rm{c}}$, which is corresponding to the minimum trap depth $U_{\rm{min}}$~\cite{C_Tuchendler_Energy_distribution}.Therefore, by measuring the probability of survival after this probability, we achieve the cumulative probability distribution over the energy of the atom. If an atom follows the Maxwell-Boltzmann distribution, a cumulative probability distribution of atom, which has lower energy than cut-off energy $E_c$, is given by \cite{C_Tuchendler_Energy_distribution}
\begin{equation}
P(E_c)= \int^{E}_{c}f_{MB}(E') dE'   =  1-[1+\eta+\frac{1}{2}\eta^2]e^{-\eta},
\label{P_MB}
\end{equation}
where $f_{\rm{MB}}$ is the normalized Maxwall-Boltzmann energy distribution function for atom and $\eta = E_c/k_BT$. 


\noindent \\
\noindent
{\bf Optical tweezer Model \Romannum{3}} \\
\noindent
The objective of model \Romannum{3} is to develop a more realistic optical tweezer model, by including unwanted fluctuations of the system, such as distributions of atomic position and velocities and fluctuation of the trap depth. The Maxwell-Boltzmann distribution describes how the distances and velocities of atoms are distributed in accordance with the temperature of the atom, as $ {\Delta}x =  \sqrt{{k_B T}/{m{\omega}_0^2}}$ and $\Delta v  =  \sqrt{{k_B T}/{m}}$.  To simulate the experimental situation, we used the Monte Carlo method to randomize the initial velocities and positions of the atoms, repeating this process 200 times. The force received by the atom from the optical tweezer is calculated in real time based on the initial settings of the atom and the path of the tweezer. Subsequently, the force received by the atom is adjusted to reflect the fluctuating trap depth ($dU \sim 0.15 {\, \rm mK}$), taking into account the power fluctuations of the laser and the efficiency changes of the AOD during the trap movement. If, in 200 instances, the distance between the atom and the trap exceeds the trap's radius ($\rm{max}(\xi_G) > d_G $), the situation is deemed to be an instance of atom escape. The probability, $P(t_f,l)$, of successfully transporting the atom is then calculated through a counting method applied to the remaining cases. The calculated probability of successful transport colored as the background of the  Fig.~\ref{Fig2}(a), and the boundary region of the probability of successful transport is approximated as 
\begin{eqnarray}\label{realEq}
l_{\rm{\Romannum{3}}}< 0.336 \frac{\sqrt{3}}{5}\frac{U_0}{md}t_f^2.
\end{eqnarray}

\noindent \\
\noindent
{\bf Atomic state change by trajectories} \\
\noindent
Atomic transport is evaluated from two perspectives: the success of the transport and the supression of unwanted heating of atoms. Both perspectives can be judged by atomic state changes during the transport process. First, the success of the transport is determined by the maximum vibrational mode of the atom, max$\left\vert \Delta n (t) \right\vert$, because if the maximum state reached by the atom during the transport is equal to or higher than the energy depth of the trap, the atom is likely to escape the trap and the transport will fail. Also, the induced atomic heating by transportation means the increase at the final atomic state. As a result, based on the final vibrational mode of the atom, $\Delta n $, we can estimate how efficient the transportation was in terms of the temperature of the atoms.

In a quantum oscillator system, the change of atomic state is dependent on the force $F$, exactly meaning acceleration $a$~\cite{Nieto_dN_1965}:
\begin{eqnarray}
{\Delta}n(t) = \frac{m  \left\vert a(\omega_0) \right\vert^2}{2\hbar \omega_0},
\end{eqnarray}
where $a(\omega_0)$ is the Fourier component of $a$ at the trap frequency $\omega_0$ for the time parameter $\tau$ ($0<\tau <t$). So, based on this equation, we can estimate the state change of the atom and compare the results of moving the atom for three different paths: (1) the constant velocity trajectory (CV, $\tilde{x}(t) = \tilde{t}$) (2) the constant jerk trajectory (CJ, $\tilde{x}(t) = 3\tilde{t}^2-2\tilde{t}^3$), and (3) the STA trajectory ($\tilde x(t) = 10 \tilde{t}^3 - 15 \tilde t^4 + 6 \tilde t^5$). 

For the CV trajectory, there will be acceleration and deceleration periods during which the atom is forced by the difference between the initial and final velocity of the atom and the trap. Since the atom should be accelerated before the atom can leave the trap ($v t < d$; $v = l/t_f$), the time for acceleration and deceleration of the atom is assumed to be $\tau_{\rm acc} \sim d/v$, and the acceleration of the atom is practically given by,
\begin{equation}
a(t)=
\begin{cases}
\frac{v^2}{d}, & (0\le t \le \tau_{\rm acc}) \\
0, & (\tau_{\rm acc} < t < t_f - \tau_{\rm acc}) \\
-\frac{v^2}{d} & (t_f-\tau_{\rm acc} \le t \le t_f). \\
\end{cases}
\end{equation}
When $\omega_0 \tau_{\rm acc} \ll 1$, the maximal $\left\vert a(\omega_0) \right\vert^2$ and the average final $\left\vert a(\omega_0) \right\vert^2$ is both about $v^2$. Therefore, the maximal atomic state change and final atomic state change is same as  max$\left\vert \Delta n (t) \right\vert$ $ = \Delta n  = m l^2/2\hbar \omega_0 t_f^2 $. For the CJ path, the acceleration of atom is $a(t) = 6l(1-2\tilde{t})/t_f^3$. When $\omega_0 t_f > 1$, the maximal $\left\vert a(\omega_0) \right\vert^2$ is about $108l/\omega_0^2t_f^4$, and the average final $\left\vert a(\omega_0) \right\vert^2$ is about $72l/\omega_0^2t_f^4$. As a result, the maximal atomic state change is max$\left\vert \Delta n (t) \right\vert = 54ml^2/\hbar\omega_0^3t_f^4$ and final atomic state change is same as  $ \Delta n  = 36ml^2/\hbar\omega_0^3t_f^4$. For the STA path, the acceleration of atom is $a(t) = 60l(\tilde{t}-3\tilde{t}^2+2\tilde{t}^3)/t_f^2$. When $\omega_0 t_f > 1$, the maximal $\left\vert a(\omega_0) \right\vert^2$ is about $100l/3\omega_0^2t_f^4$, and the average final $\left\vert a(\omega_0) \right\vert^2$ is about $7200l^2/\omega_0^4t_f^6$. As a result, the maximal atomic state change is max$\left\vert \Delta n (t) \right\vert \simeq 50ml^2/3\hbar\omega_0^3t_f^4$ and final atomic state change is same as  $ \Delta n  = 3600ml^2/\hbar\omega_0^5t_f^6$.

\end{document}